\documentclass{tlp}
\title[O(1) navigation and editing]
      {O(1) reversible tree navigation without cycles}
\author[R. A. O'Keefe]
       {RICHARD A. O'KEEFE\\
	The University of Otago, NZ\\
	\email{ok@cs.otago.ac.nz}}
\begin{document}

\pagerange{\pageref{firstpage}--\pageref{lastpage}}
\volume{\textbf{1} (1):}
\jdate{January 2001}
\setcounter{page}{1}
\pubyear{2001}

\newcommand\othrpearl{P\ls R\ls O\ls G\ls R\ls A\ls M\ls M\ls I\ls  N\ls G\ns P\ls E\ls A\ls R\ls L}

\maketitle[o]

\shorttitle{Programming pearl}

\label{firstpage}

\begin{abstract}
Imperative programmers often use cyclically linked trees
in order to achieve O(1) navigation time to neighbours.
Some logic programmers believe that cyclic terms are necessary
to achieve the same in logic-based languages.
An old but little-known technique
provides O(1) time and space navigation without cyclic links,
in the form of reversible predicates.
A small modification provides O(1) amortised time and space editing.
\end{abstract}

\section{The problem}

When imperative programmers think of lists, they commonly choose
doubly linked lists, instead of the singly linked lists that logic
and functional programmers use.  In the same way, it is extremely
common for trees to be given parent links, whether they are really
needed or not.  A typical C example might be
\begin{verbatim}
typedef int Datum;

typedef struct TreeRec *TreePtr;
struct TreeRec {
    TreePtr left, right, up, down;
    Datum datum;
};
\end{verbatim}
where `down' points to the first child of a node, `up' to its parents,
and the children of a node form a doubly linked list with `left' and
`right' pointers.  Essentially this representation is required by the
Document Object Model \cite{dom1,dom2}, for example.

Cyclically linked trees in imperative languages such as Java provide
constant time navigation in any of the four directions (up, down, left,
right) and also constant time and constant space editing (insert, delete,
replace).
They do so at a price:  each element is rigidly locked into place, so that
any kind of space sharing (such as hash consing) is made impossible.

Some logic programming languages have been designed to support
cyclically linked terms.  That does provide constant time navigation,
but not editing.  The locking into place that is a nuisance in
imperative languages is a very serious difficulty in logic programming
languages.  Additionally, reasoning about the termination of programs
that traverse cyclic graphs is harder than reasoning about programs that
traverse trees, whether in Prolog dialects or in lazy functional
languages, so it is useful to get by with trees if we can.

This article has two parts.  In the first part, I present ``fat pointers''
that can be used to navigate around pure trees.  The tree itself remains
unmodified throughout.  The main predicates I define have the form
from\_to(From,To).  If one of the arguments is ground and the other is
uninstantiated, the time and space cost is O(1) per solution.  

In the second part, I present ``edit pointers'' that can be used to
navigate around trees and edit them, in O(1) amortised time and space
per step and edit.

The type declarations are Mycroft/O'Keefe \cite{mycroft}
type declarations using the
syntax of Mercury \cite{mercury}.  The predicate declarations are also
Mycroft/O'Keefe declarations giving argument types and modes.  The code
has been type-checked by the Mercury compiler.  The clauses are
Edinburgh Prolog.  This paper provides evidence that using different
definitions for different modes is useful, but that is difficult in
Mercury, so the modes were not converted to Mercury syntax and the code
does not pass Mercury mode-checking.

\section{Acknowledgement}

This is a generalisation of a method for O(1) left and right navigation
in a list shown to me by David H.~D.~Warren in 1983, in a text editor he
wrote in Prolog.

A companion paper \cite{domcurse} presents this technique in a
functional context.  It was rejected on the grounds that the data
structure had already been published by Huet as the Zipper in
\cite{zipper}.  However, the two data structures described in this
paper and in \cite{domcurse} are different from the Zipper, and
the issues discussed here are different.

\section{Fat Pointers}

The key idea is to distinguish between a tree and a pointer into a tree.
The usual C/Java approach blurs this distinction, and that has misled some
logic programmers into thinking that cyclically linked trees are
necessary in order to obtain a certain effect in pointers.
A tree just has to represent certain information; a pointer has
to know how to move.

A suitable data type for trees is
\begin{verbatim}
:- type tree(Datum)
   ---> node(Datum, list(tree(Datum))).

:- pred tree_datum(tree(D), D).
tree_datum(node(Datum,_), Datum).

:- pred tree_children(tree(D), list(tree(D))).
tree_children(node(_,Children), Children).
\end{verbatim}

Like a C pointer, a ``fat pointer'' points to a specific (sub)tree;
unlike a C pointer, a ``fat pointer'' carries a context:  the
(sub)tree's left siblings (ordered from nearest to furthest), its right
siblings (ordered from nearest to furthest), and a parent fat pointer,
if this is not the root.
\begin{verbatim}
:- type pointer(D)
   ---> ptr(tree(D), list(tree(D)), list(tree(D)), pointer(D))
      ; no_ptr.
\end{verbatim}

\noindent The predicates we define will never be true of a `no\_ptr' argument.

\begin{verbatim}
:- pred top_pointer(tree(D), pointer(D)).
top_pointer(Tree, ptr(Tree,[],[],no_ptr)).

:- pred pointer_tree(pointer(D), tree(D)).
pointer_tree(ptr(Tree,_,_,_), Tree).

:- pred pointer_datum(pointer(D), D).
pointer_datum(ptr(Tree,_,_,_), Datum) :-
    tree_datum(Tree, Datum).

:- pred at_left(pointer(_)).
at_left(ptr(_,[],_,_)).

:- pred at_right(pointer(_)).
at_right(ptr(_,_,[],_)).

:- pred at_top(pointer(_)).
at_top(ptr(_,_,_,no_ptr)).

:- pred at_bottom(pointer(_)).
at_bottom(ptr(Tree,_,_,_)) :-
    tree_children(Tree, []).

:- pred left_right(pointer(D), pointer(D)).
left_right(ptr(T,L,[N|R],A), ptr(N,[T|L],R,A)).

:- pred up_down_first(pointer(D), pointer(D)).
up_down_first(P, ptr(T,[],R,P)) :-
  % P = ptr(tree(_,[T|R]),_,_,_).
    P = ptr(Tree,_,_,_),
    tree_children(Tree, [T|R]).
\end{verbatim}

The `top\_pointer/2' predicate may be used to make a fat pointer
from a tree, or to extract a tree from a fat pointer positioned at\_top.
The `at\_*/1' predicates recognise whether a fat pointer is positioned
at a boundary.
The query `left\_right(Left, Right)' is true when Left and Right are
pointers to adjacent siblings, Left on the left, and Right on the right.
The query `up\_down\_first(Up, Down)' is true when Up is a pointer and
Down is a pointer to Up's first child; it is O(1) time and space in either
direction provided that the list of children of a node is stored in that
node and not recomputed.

The query `up\_down(Up, Down)' is to be true when Up is a pointer and
Down is a pointer to any of Up's children.  It uses mode-dependent code.
\begin{verbatim}
:- pred up_down(pointer(D), pointer(D)).
up_down(P, ptr(T,L,R,A)) :-
    (   var(P) ->
        A = ptr(_,_,_,_), % not no_ptr, that is.
        P = A
    ;   A = P,
        P = ptr(Tree,_,_,_),
        tree_children(Tree, Children),
        % split Children++[] into reverse(L)++[T]++R
        split_children(Children, [], L, T, R)
    ).

:- pred split_children(list(T), list(T), list(T), T, list(T)).
split_children([T|R], L, L, T, R).
split_children([X|S], L0, L, T, R) :-
    split_children(S, [X|L0], L, T, R).
\end{verbatim}

We could write this clearly enough as
\begin{verbatim}
up_down_specification(P, ptr(T,L,R,P)) :-
    P = ptr(tree(_,Children),_,_,_),
    split_children(Children, [], L, T, R).
\end{verbatim}
but then the cost of moving up would not be O(1).  This is an
interesting predicate, because getting efficient behaviour in multiple
modes is not just a matter of subgoal ordering.  In the (+,-) mode, it
is acceptable to call `split\_children/5', because that will be O(1)
space and time per solution.  In the (-,+) mode, we must bypass that
call.  If the code is encapsulated in a module and type checked, it is
clear that the bypassed call must succeed, but that would not be obvious
to a compiler.

Queries in SGML and XML often ask for (not necessarily adjacent)
preceding or following siblings of a node.  Testing whether one node
is a following sibling of another does not require mode-dependent
code.  However, in order to prevent unbounded backtracking in the
reverse direction, the code below uses the technique of passing a
variable (L2) twice:  once for its value, and once so that its length
can act as a depth bound.

\begin{verbatim}
:- pred siblings_before_after(pointer(D), pointer(D)).
siblings_before_after(ptr(T1,L1,R1,A), ptr(T2,L2,R2,A)) :-
    right_move([T1|L1], R1, L2, [T2|R2], L2).

:- pred right_move(list(T), list(T), list(T), list(T), list(T)).
right_move(L, R, L, R, _).
right_move(L1, [X|R1], L2, R2, [_|B]) :-
    right_move([X|L1], R1, L2, R2, B).
\end{verbatim}
%

Moving from left to right, this costs $O(1)$ per following sibling.
Moving from right to left, it costs $O(n)$ per preceding sibling
because the second clause builds up patterns of $O(n)$ average length
and the first clause unifies them against the known lists for each
solution.

The `left\_right\_star/2' predicate below fixes this using
mode-dependent code.  I do not know whether $O(1)$ cost per solution
can be obtained with bidirectional code.

Next we have transitive (\_plus) and reflexive transitive (\_star)
closures of the basic predicates.  Once again, we want to execute
different code for the (+,-) and (-,+) modes of the `*\_plus/2' predicates.
\begin{verbatim}
:- pred left_right_star(pointer(D), pointer(D)).
left_right_star(L, R) :-
    (   L = R
    ;   left_right_plus(L, R)
    ).

:- pred left_right_plus(pointer(D), pointer(D)).
left_right_plus(L, R) :-
    (   var(L) ->
        left_right(X, R),
        left_right_star(L, X)
    ;   left_right(L, X),
        left_right_star(X, R)
    ).

:- pred up_down_star(pointer(D), pointer(D)).
up_down_star(A, D) :-
    (   A = D
    ;   up_down_plus(A, D)
    ).

:- pred up_down_plus(pointer(D), pointer(D)).
up_down_plus(A, D) :-
    (   var(A) ->
        up_down(X, D),
        up_down_star(A, X)
    ;   up_down(A, X),
        up_down_star(X, D)
    ).
\end{verbatim}

\section{Performance}

%
%

A simple benchmark is to build a large tree and traverse it.
\begin{verbatim}
:- pred run.
run :-
    mk_tree(10, T),
    time((direct_datum(T, X), atom(X))),
    time((any_pointer_datum(T, X), atom(X))),    
    time(labels(T,_)),
    time(collect(T,_)).

:- pred mk_tree(integer, tree(integer)).
mk_tree(D, node(D,C)) :-
    (   D > 0 ->
        D1 is D - 1,
        C = [T,T,T,T],
        mk_tree(D1, T)
    ;   C = []
    ).

:- pred direct_datum(tree(D), D).
direct_datum(node(D,_), D).
direct_datum(node(_,C), D) :-
    member(N, C),
    direct_datum(N, D).

:- pred any_pointer_datum(tree(D), D).
any_pointer_datum(T, D) :-
    top_pointer(T, P),
    up_down_star(P, N),
    pointer_datum(N, D).

:- pred labels(tree(D), list(D)).
labels(Tree, Labels) :-
    labels(Tree, Labels, []).

:- pred labels(tree(D), list(D), list(D)).
labels(node(Label,Children)) -->
    [Label], labels_list(Children).

:- pred labels_list(list(tree(D)), list(D), list(D)).
labels_list([]) --> [].
labels_list([Tree|Trees]) -->
    labels(Tree), labels_list(Trees).

:- pred collect(tree(D), list(D)).
collect(Tree, Labels) :-
    top_pointer(Tree, Ptr),
    collect(Ptr, Labels, []).

:- pred collect(pointer(D), list(D), list(D)).
collect(Ptr, [Datum|Labels1], Labels) :-
    pointer_datum(Ptr, Datum),
    (   at_bottom(Ptr) -> Labels1 = Labels2
    ;   up_down_first(Ptr, Child),
        collect(Child, Labels1, Labels2)
    ),
    (   at_right(Ptr) -> Labels2 = Labels
    ;   left_right(Ptr, Sibling),
        collect(Sibling, Labels2, Labels)
    ).
\end{verbatim}
this builds a tree with 1,398,101 nodes,
and traverses it using a backtracking search (both directly and using
``fat pointers''), and by building a list (both directly and using
``fat pointers'').

\begin{table}
\caption{Traversal times in seconds, by methods and languages}
\label{traversal-table}
\begin{minipage}{\textwidth}
\begin{tabular}{rrrrl}
\hline\hline
search & search  & list   & list  & \\
direct & pointer & direct & pointer & system\\\hline
\hline
 89.5s &  235.0s & 151.2s &  197.0s & SWI Prolog 3.3.8\\
 13.2s &   41.6s &  10.5s &   44.7s & Quintus Prolog 3.4\\
  7.0s &   16.9s &   3.9s &   22.5s & SICStus Prolog 3.8.4\\\hline
  N/A  &    N/A  &   8.0s &   17.0s & ghc 4.08.1\\
  N/A  &    N/A  &   2.5s &    5.6s & clm 1.3.1\\
\hline\hline
\end{tabular}
\end{minipage}
\end{table}

Table \ref{traversal-table} shows the times for this benchmark, as
measured on an 84~MHz SPARCstation 5.  The last two rows of table
\ref{traversal-table} refer to lazy functional languages:  Haskell
(ghc), and Clean (clm).  Both of those compilers use whole-program
analysis, inlining, and strong type information, unlike the incremental,
direct, and untyped Prolog compilers.

\section{The findall/3 sharing problem}

The first draft of this article included measurements for findall/3.
They were horrifying.  This data structure is bad news for findall/3
and its relatives.  To understand why, let's consider three queries:

\begin{verbatim}
q1(P, L) :- findall(Q, p1(P, Q), L).
q2(P, L) :- findall(T, p2(P, T), L).
q3(P, L) :- findall(D, p3(P, D), L).

p1(P, Q) :- up_down_star(P, Q).
p2(P, T) :- p1(P, Q), Q = ptr(T, _).
p3(P, D) :- p1(P, Q), pointer_datum(Q, D).
\end{verbatim}

For concreteness, suppose that P points to a complete binary tree with
$n 
$ nodes,
and that the node data are one-word constants.

Query q3 requires 
$O(n)$ space to hold the result.

Query q2 requires 
$O(n\log n)$ space to hold the result.
This is because fragments of the tree are repeatedly copied.

Query q1 requires at least 
$O(n^2)$ space to hold the result.
Every pointer holds the entire tree to simplify movement upwards,
so the entire tree and its fragments are repeatedly copied.

If findall/3 copied terms using some kind of hash consing,
the space cost could be reduced to $O(n)$, but not the time
cost, because it would still be necessary to test whether a
newly generated solution could share structure with existing
ones.

One referee suggested that Mercury's `solutions/2'
would be cleverer.  A test in the 0.10 release showed that 
it is not yet clever enough.

This is related to the problem of costly pointer unification:
a fat pointer contains the entire original tree and many of its
fragments as well as the fragment it directly refers to.  But
that is only a problem when you try to copy a fat pointer or
unify it with something.

The `siblings\_before\_after/2' predicate is efficient if one of the
arguments is ground and the other uninstantiated.  With two ground
arguments, it is expensive, because
because testing whether two fat pointers are the same fat pointer
is expensive in this representation.  Two fat pointers are the same if and
only if they unify, but that may involve comparing the source tree with
itself (and its subtrees with themselves, perhaps repeatedly).

After reading an earlier draft of this, Thomas Conway \cite{conway}
suggested a variant in which unifying fat pointers would be $O(n)$
instead of $O(n^2)$.  I have not adopted his code, because it makes
moving up take more time and turn over more memory.  He stores just
enough of a parent to reconstruct it, given the child, instead of the
entire parent.  Conway's code is very nearly Huet's Zipper \cite{zipper}.

\section{Sets of Pointers}

Thomas Conway \cite{conway} has pointed out that it would be useful
to form sets of (pointers to) nodes, and that fat pointers are not
suitable for that because they are so costly to unify.
Sets of nodes are required for the HyTime query language and by other
query languages, including XPath \cite{xpath}.

When one wants to form a set of subtrees, one might want a set of
tree values, or a set of occurrences of tree values.

Using a pointer forces you to work with occurrences rather than
values, and that is not always appropriate.  The comparison of
subtrees can be made more efficient by preprocessing the main tree
so that every node contains a hash code calculated from its value.

Preprocessing the main tree so that every node has a unique number
would make pointer comparison fast too.  The `pair' type here is
standard in Mercury, tracing back to `keysort/2' in DEC-10 Prolog.

\begin{verbatim}
:- type ntree(D)    = tree(pair(integer,D)).
:- type npointer(D) = pointer(pair(integer,D)).

:- pred number_tree(tree(D), ntree(D)).
number_tree(Tree, Numbered) :-
    number_tree(Tree, Numbered, 0, _).

:- pred number_tree(tree(D), ntree(D), integer, integer).
number_tree(node(Datum,Children),
            node(N0-Datum,NumberedChildren), N0, N) :-
    N1 is 1 + N0,
    number_children(Children, NumberedChildren, N1, N).

:- pred number_children(list(tree(D)), list(ntree(D)),
                        integer, integer).
number_children([], [], N, N).
number_children([C|Cs], [D|Ds], N0, N) :-
    number_tree(C, D, N0, N1),
    number_children(Cs, Ds, N1, N).

:- pred same_pointer(npointer(D), npointer(D)).
same_pointer(P1, P2) :-
    pointer_datum(P1, N-_),
    pointer_datum(P2, N-_).

:- pred compare_pointers(order, npointer(D), npointer(D)).
compare_pointers(O, P1, P2) :-
    pointer_datum(P1, N1-_),
    pointer_datum(P2, N2-_),
    compare(O, N1, N2).
\end{verbatim}

Prolog, having no type classes, would require special purpose set
operations for sets of fat pointers using this technique.
Mercury, having type classes, would not require special code.

\section{Hello Heraclitus, Goodbye Parmenides}

The data structure described in the previous section implements pointers
that point at particular nodes in trees, may exist in large numbers,
and may be stepped in any direction at O(1) cost.  It is useful for
navigating around trees that are not to be changed.

If you are writing an editor, it is a blunder to try to point
\emph{at} nodes.
Instead, it is advisable to point \emph{between} nodes, as Emacs does.
We need a different data structure, and different operations.

The real challenge, as Huet understood well in \cite{zipper}, is to show
that declarative languages can handle editing effectively.

The title of this article promised $O(1)$ reversible navigation; it did
not promise $O(1)$ reversible editing.  Reversible editing is attainable,
and so is $O(1)$ editing.  My attempts to discover a data structure that
supports $O(1)$ reversible editing have so far failed.

Because these operations are not reversible, I show input (+) and output (-)
modes in the :- pred\cite{zipper}  declarations.

The normal way in Prolog to ``change'' a node at depth $d$ in a tree
is to build a new tree sharing as much as possible with the old one;
that takes $O(d)$ time and space.  The trick that permits $O(1)$ cost
edit operations is not to rebuild the new tree at once, but to interleave
rebuilding with editing and navigation.  An \verb|edit_ptr| records the
subtrees to either side of the current position, whether the sequence at
this level has been edited, and what the parent position was.

To ``change'' trees, we need a new operation that creates a new node
with the same information as an old one, except for a new list of children.
\begin{verbatim}
:- pred tree_rebuild(+list(tree(D)), +tree(D), -tree(D)).
tree_rebuild(C, node(D,_), node(D,C)).
\end{verbatim}

The `extract/2' predicate extracts the edited Tree from an \verb|edit_ptr|.

Since these predicates are not reversible, mode information is shown
in the :-pred declarations.
\begin{verbatim}
:- type nlr(P)
   ---> no_parent         % this is the root
      ; left_parent(P)    % came down on the left
      ; right_parent(P).  % came down on the right.

:- type changed
   ---> n
      ; y.

:- type edit_ptr(D)
   ---> edit_ptr(list(tree(D)), list(tree(D)),
                 changed, nlr(edit_ptr(D))).
\end{verbatim}

An edit pointer contains a list of the nodes preceding the current
position (in near-to-far order), a list of the nodes following the
current position (in near-to-far order), a `changed' flag, and a
parent pointer tagged so that we know whether we came down on the
left or the right.

One of the differences between this data structure and the Zipper
\cite{zipper} is that the Zipper builds new retained structure as
it moves, while this data structure uses a `changed' flag to
revert to the original structure if nothing has happened but movement.

\begin{verbatim}
:- pred start(+tree(D), -edit_ptr(D)).
start(T, edit_ptr([],[T],n,no_parent)).

:- pred at_top(+edit_ptr(_)).
at_top(edit_ptr(_,_,_,no_parent)).

:- pred at_left(+edit_ptr(_)).
at_left(edit_ptr([],   _,_,_)).

:- pred left_datum(+edit_ptr(D), ?D).
left_datum(edit_ptr([T|_],_,_,_), D) :- tree_datum(T, D).

:- pred left(+edit_ptr(D), -edit_ptr(D)).
left(edit_ptr([T|L],R,C,P), edit_ptr(L,[T|R],C,P)).

:- pred left_insert(+tree(D), +edit_ptr(D), -edit_ptr(D)).
left_insert(T, edit_ptr(L,R,_,P), edit_ptr([T|L],R,y,P)).

:- pred left_delete(-tree(D), +edit_ptr(D), -edit_ptr(D)).
left_delete(T, edit_ptr([T|L],R,_,P), edit_ptr(L,R,y,P)).

:- pred left_replace(+tree(D), +edit_ptr(D), -edit_ptr(D)).
left_replace(T, edit_ptr([_|L],R,_,P), edit_ptr([T|L],R,y,P)).

:- pred left_down(+edit_ptr(D), -edit_ptr(D)).
left_down(P, edit_ptr([],K,n,left_parent(P))) :-
    P = edit_ptr([T|_],_,_,_),
    tree_children(T, K).

:- pred left_promote_children(+edit_ptr(D), -edit_ptr(D)).
left_promote_children(edit_ptr([T|L],R,_,P),
                      edit_ptr(L1,R,y,P)) :-
    tree_children(T, K),
    reverse_append(K, L, L1).

:- pred reverse_append(+list(T), +list(T), -list(T)).
reverse_append([], R, R).
reverse_append([X|S], R0, R) :-
    reverse_append(S, [X|R0], R).

:- pred at_right(+edit_ptr(_)).
at_right(edit_ptr(_,[],_,_)).

:- pred right_datum(+edit_ptr(D), -D).
right_datum(edit_ptr(_,[T|_],_,_), D) :- tree_datum(T, D).

:- pred right(+edit_ptr(D), -edit_ptr(D)).
right(edit_ptr(L,[T|R],C,P), edit_ptr([T|L],R,C,P)).

:- pred right_insert(+tree(D), +edit_ptr(D), -edit_ptr(D)).
right_insert(T, edit_ptr(L,R,_,P), edit_ptr(L,[T|R],y,P)).

:- pred right_delete(-tree(D), +edit_ptr(D), -edit_ptr(D)).
right_delete(T, edit_ptr(L,[T|R],_,P), edit_ptr(L,R,y,P)).

:- pred right_replace(+tree(D), +edit_ptr(D), -edit_ptr(D)).
right_replace(T, edit_ptr(L,[_|R],_,P), edit_ptr(L,[T|R],y,P)).

:- pred right_down(+edit_ptr(D), -edit_ptr(D)).
right_down(P, edit_ptr([],K,n,right_parent(P))) :-
    P = edit_ptr(_,[T|_],_,_),
    tree_children(T, K).

:- pred right_promote_children(+edit_ptr(D), -edit_ptr(D)).
right_promote_children(edit_ptr(L,[T|R],_,P),
                       edit_ptr(L,R1,y,P)) :-
    tree_children(T, K),
    append(K, R, R1).

:- pred up(+edit_ptr(D), -edit_ptr(D)).
up(edit_ptr(_,_,n,left_parent(P)), P).
up(edit_ptr(_,_,n,right_parent(P)), P).
up(edit_ptr(X,Y,y,left_parent(edit_ptr([T|L],R,_,P))),
   edit_ptr([N|L],R,y,P)) :-
    reverse_append(X, Y, K),
    tree_rebuild(K, T, N).
up(edit_ptr(X,Y,y,right_parent(edit_ptr(L,[T|R],_,P))),
   edit_ptr(L,[N|R],y,P)) :-
    reverse_append(X, Y, K),
    tree_rebuild(K, T, N).

:- pred extract(+edit_ptr(D), -tree(D)).
extract(edit_ptr(_,[T|_],_,no_parent), Tree) :- !, Tree = T.
extract(P, Tree) :- up(P, P1), extract(P1, Tree).
\end{verbatim}

It is easy to see that each of these operations except `up' and
`extract' requires $O(1)$ time and $O(1)$ space, assuming that
`tree\_children' is $O(1)$.  Moving `up' from a position with $n$
nodes to its left takes $O(n)$ time and space, but that position must
have been reached by a minimum of $n$ `right/2' and/or `left\_insert/3'
operations, so in the single-threaded case all the basic operations are
$O(1)$ amortised time and space.

\section{Why isn't editing reversible?}

Essentially, because the editing operations destroy information.
We can easily reverse the operations ``move right'' or ``move down'',
because the only thing that has changed is the position, and the
name of the predicate tells us what the old position must have been.

Suppose, however, we tried to make insertion and deletion the same
operation:
\begin{verbatim}
:- pred indel_right(D, edit_ptr(D), edit_ptr(D)).
indel_right(T, edit_ptr(L,R,Fs,P), edit_ptr(L,[T|R],Fw,P)) :-
    ?
\end{verbatim}
We would not know how to define Fs and Fw.

The edit\_pointer data structure may not have reversible operations, but
it is a persistent\footnote{``Persistence'' has at least two meanings:
out-living a program execution (as in PS-Algol) or surviving changes so
that the original value can be resurrected (as here).}  data structure,
so that an editor using it can easily support unbounded UNDO.

\section{An example}

Imagine an application to HTML, where we might have
\begin{verbatim}
:- type html_info
   ---> element(atom,list(pair(atom,string)))
      ; pcdata(string).
:- type html_tree == tree(html_info).
\end{verbatim}
The \verb|<FONT>| tag is almost always misused, and is not allowed in
``strict'' HTML4 or XHTML.  So we might want to replace every
\verb|<FONT>| element by its contents.

\begin{verbatim}
:- pred unfont(+html_tree, -html_tree).
unfont(HTML0, HTML) :-
    start(HTML0, Ptr0),
    unfont_loop(Ptr0, Ptr),
    extract(Ptr, HTML).

:- pred unfont_loop(+edit_ptr(html_info), -edit_ptr(html_info)).
unfont_loop(P0, P) :-
    (   at_right(P0) -> P = P0
    ;   right_datum(P0, element(font,_)) ->
        right_promote_children(P0, P1),
        unfont_loop(P1, P)
    ;   right_down(P0, P1),
        unfont_loop(P1, P2),        
        up(P2, P3),
        right(P3, P4),
        unfont_loop(P4, P)
    ).
\end{verbatim}

\section{Conclusion}

We don't need cyclic links or mutable objects in declarative languages
to support efficient multiway traversal and editing of trees, including
models of XML.

\end{document}